\documentclass[conference]{IEEEtran}
\IEEEoverridecommandlockouts
\usepackage{cite}
\usepackage{url}
\usepackage{amsmath,amssymb,amsfonts}
\usepackage{algorithmic}
\usepackage{graphicx}
\usepackage{textcomp}
\usepackage{xcolor}
\usepackage[linesnumbered,ruled,vlined]{algorithm2e}
\def\BibTeX{{\rm B\kern-.05em{\sc i\kern-.025em b}\kern-.08em
    T\kern-.1667em\lower.7ex\hbox{E}\kern-.125emX}}
\begin{document}

\title{Mitigating Challenges in Ethereum's Proof-of-Stake Consensus: Evaluating the Impact of EigenLayer and Lido\\
}

\author{\IEEEauthorblockN{Li Li}
\textit{Virginia Tech}\\
}

\maketitle

\begin{abstract}
The transition of Ethereum from a Proof-of-Work (PoW) to a Proof-of-Stake (PoS) consensus mechanism introduces a transformative approach to blockchain validation, offering enhanced scalability, energy efficiency, and security. However, this shift also presents significant challenges, including high barriers to becoming a validator, restrictions on the liquidity of staked Ether (ETH), and the risk of centralization due to staking pool dynamics. This paper addresses these challenges by exploring two innovative solutions: EigenLayer and Lido.

EigenLayer is a middleware solution enabling restaking, allowing validators to secure multiple protocols and thereby increasing decentralization and profitability. Lido, a liquid staking protocol, simplifies participation by issuing stETH tokens that retain liquidity, allowing users to earn rewards without long-term lock-up constraints. This paper provides a detailed analysis of how these technologies mitigate key PoS challenges, reduce validator entry barriers, unlock staked capital, and improve decentralization. We conclude with an evaluation of the combined potential of EigenLayer and Lido to foster a more resilient and inclusive Ethereum ecosystem, setting the stage for further advancements in decentralized finance.

\end{abstract}


\section{Introduction}
Proof-of-Work (PoW)\cite{Bitcoin_Wiki_PoW} consensus mechanisms have long been the cornerstone of validation and transaction execution since the invention of the on-chain cryptocurrency, Bitcoin. These systems rely on miners' computational power to secure and maintain the blockchain. However, PoW systems are not without their drawbacks, most notably their substantial impact on energy consumption and environment. As the cryptocurrency space continues to grow and evolve, there is increasingly more interest in transitioning towards more energy-efficient or "green" consensus mechanisms, with Proof-of-Stake (PoS)\cite{Bitcoin_Wiki_PoS} emerging as a promising alternative.

PoS systems introduce a fundamental shift in how block validation and authentication occurs on the blockchain. Unlike PoW, where miners compete with other miners to solve complex cryptographic puzzles, PoS relies on the number of cryptocurrency tokens held by a node or a validator, determining its ability and trustworthiness to validate transactions and create new blocks. This shift not only reduces energy consumption but also offers the possibility of broader participation in network validation. Validators in a PoS system are selected arbitrarily, and their ability to create and authenticate blocks is directly proportional to number of tokens they hold and "stake" as the collateral.

Within the realm of PoS-based initiatives, the conventional approach to staking involves individuals locking up their tokens within a single project for an extended timed period. In return, as an incentive, they anticipate receiving predetermined staking rewards. However, this staking model presents a significant barrier to entry for many potential participants, as it often requires substantial initial investment.

In response to this challenge, innovative solutions have emerged, including staking pools and re-staking middleware. These solutions enables a third-party entity to act as a validator operator on behalf of individual investors, who participate in staking activities within the cryptocurrency network. By pooling their resources with others, participants can enjoy the benefits of staking without the need for significant capital outlays.

One prominent blockchain network that has embraced PoS and staking is Ethereum\cite{ethereum-whitepaper-2014}. Ethereum's transition to Ethereum 2.0, often referred to as ETH2, is a multi-phased upgrade aimed at addressing scalability, security, and energy efficiency. Ethereum 2.0 introduces a pivotal change by migrating from PoW to PoS, where validators lock up a specified amount of Ether (ETH) as collateral to participate in block validation and consensus. This shift will significantly reduce Ethereum's energy consumption, making it more sustainable and scalable.

\subsection{What are the Challenges of Ethereum PoS?}

\subsubsection{High barrier to become a validator}
Validators play a crucial role in securing the Ethereum network and validating transactions. They are responsible for validating and confirming transactions on the blockchain, maintaining network security, consensus, and ensuring the integrity of the Ethereum ecosystem. Validators are part of the proof-of-stake (PoS) consensus mechanism, which Ethereum plans to transition to with the upcoming Ethereum 2.0 upgrade. Unlike proof-of-work (PoW) systems that rely on miners to solve complex mathematical puzzles, PoS systems determine validators based on the number of coins they hold and are willing to “stake” as collateral. Validators are randomly selected to propose blocks and validate transactions by locking up a specified amount of Ether (ETH) as a security deposit. Once selected, validators add new blocks to the blockchain and validate transactions by ensuring they meet the network’s rules and consensus algorithms. Validators play a critical role in maintaining the integrity and security of the Ethereum network.\cite{HowToBecomeValidator}

However, becoming a validator on Ethereum requires a significant commitment of time, resources, and technical knowledge 1. It’s not a task to be taken lightly, but with dedication and the right approach, it can be a rewarding endeavor 1. The first step towards becoming a validator on Ethereum is to ensure you have a sufficient amount of Ether (ETH) to meet the staking requirements. At present, the Ethereum 2.0 Beacon Chain requires a minimum stake of 32 ETH to become a validator.

\subsubsection{Staked ETH is locked and cannot be withdrawn or used}
Staked ETH is a critical component of Ethereum’s proof-of-stake consensus mechanism, which is used to secure the network and process transactions. However, staked ETH is locked and cannot be withdrawn or used until the validator exits the active validator set. This means that stakers must commit to a long-term lockup period, which can last up to several years.

This lockup period presents several challenges for Ethereum, including centralization, censorship, and exploitation from certain infrastructure intermediaries. For instance, stakers may be unable to access their assets during the lockup period, which can lead to potential losses. Additionally, the inability to withdraw staked ETH can limit the liquidity of the asset, which can make it difficult to use in other applications.

\subsubsection{Small number of entities could control a large majority of the staked ETH}

Ethereum, the largest smart contract blockchain, switched to proof-of-stake (PoS) from proof-of-work (PoW) in 2022. Under PoS, Ethereum validators stake 32 ether on the network, and they are randomly selected to add blocks. With staking, Ethereum drastically cut the blockchain’s environmental impact, but it continues to face a slew of challenges around centralized power, censorship, and exploitation from certain infrastructure intermediaries.

\begin{figure}[!t]
    \centering
    \includegraphics[width=0.5\textwidth]{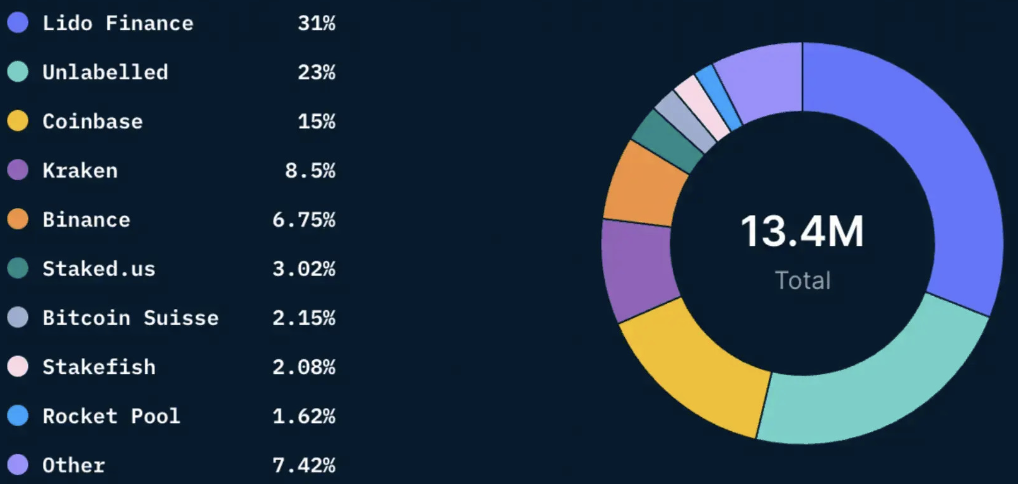}
    \caption{Portion of Staked ETH.}
    \label{fig:PortionETH}
\end{figure}

One of the dangers for Eth 2.0 as a result of its staking dynamics is that a single large holder of ether – be it a cryptocurrency whale, exchange or staking pool – could monopolize control over the majority of active validators in the network. We can see the figure \ref{fig:PortionETH}, it shows that a small amount of entities control a large amount of the staked ETH. This could lead to a situation where a small number of entities could control a large majority of the staked ETH, which could potentially compromise the security of the network.\cite{ValidPointsEthereumChallenge} 

\begin{figure}[!t]
    \centering
    \includegraphics[width=0.5\textwidth]{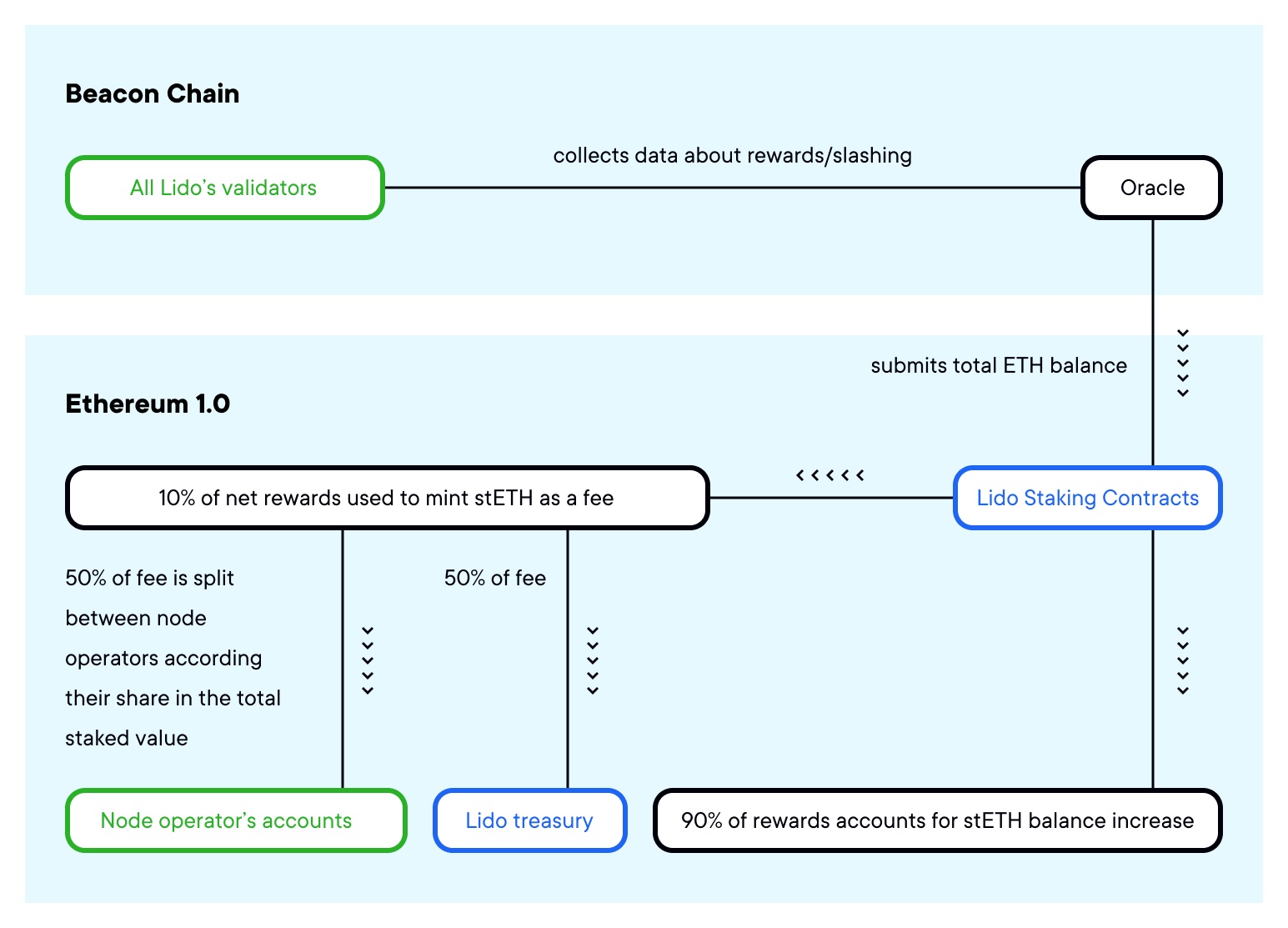}
    \caption{Oracle balance tracking process.}
    \label{fig:yourlabel}
\end{figure}

\subsection{What is Eigenlayer?}

Eigenlayer\cite{EigenLayer_Whitepaper} is a middleware layer that is designed to address some of the challenges of Ethereum PoS. It allows validators to stake their ETH on multiple protocols, which can help to improve the decentralization of the Ethereum network and reduce the risk of centralization.

Eigenlayer is not a protocol in the traditional sense, but it is a powerful tool that can be used to build new protocols and applications on top of Ethereum.

\subsection{What is Lido?}

Lido\cite{Lido:Ethereum-Liquid-Staking} is a liquid staking protocol that allows users to stake their ETH without having to lock it up. This makes it easier for users to participate in staking and earn rewards.

Lido works by pooling ETH from users and then staking it on behalf of the pool. Users are then issued a token called staked ETH (stETH) which represents their share of the pool. stETH can be traded or used in DeFi just like regular ETH.

\subsection{How can Eigenlayer and Lido Address the Challenges of Ethereum PoS?}

Eigenlayer and Lido can address the challenges of Ethereum PoS in a number of ways. Eigenlayer can help to improve the decentralization of the Ethereum network by allowing validators to stake their ETH on multiple protocol. This can reduce the risk of centralization and make the network more secure.

Lido can make it easier for users to participate in staking and earn rewards by allowing them to stake their ETH without having to lock it up. This can help to increase the number of validators on the network and further improve its security.

In addition, Eigenlayer and Lido can work together to provide users with even more flexibility and options for staking their ETH. For example, users could stake their ETH on Eigenlayer and then use Lido to liquidate their staked ETH. This would allow users to earn staking rewards while still having access to their ETH.

\section{Related Work}

Proof of Work (PoW) is a widely used consensus mechanism, initially introduced by Bitcoin, to ensure data integrity and security in decentralized networks. The core idea behind PoW is to have participants (miners) compete to solve complex mathematical puzzles through hash computations, with the winner earning the right to add a new block to the blockchain and receive rewards. This mechanism not only prevents data tampering but also raises the cost for malicious actors, making attacks economically unfeasible.

The security of a PoW network heavily depends on two key factors: the distribution of computational power (hashrate) \cite{1,2,3,4} and the stability of the underlying network infrastructure\cite{5,6,7,8,9,10,wu,11}. A more distributed hashrate enhances the system’s security by mitigating the risk of a single entity gaining majority control, which would enable a "51\% attack." In addition, the stability and low latency of the network infrastructure are crucial for effective communication between nodes. If network delays are too high or if the network suffers from distributed denial-of-service (DDoS) attacks, forks can occur, undermining the network's consensus and stability.

PoW-based networks also provide strong resistance against Sybil attacks, as acquiring the necessary computational power to influence the network is prohibitively expensive. However, the high energy consumption of PoW has sparked debates about its sustainability, leading networks like Ethereum to transition to more energy-efficient consensus mechanisms, such as Proof of Stake (PoS).

Ethereum's transition to a Proof-of-Stake (PoS) consensus mechanism has been a major topic of interest, addressing challenges such as scalability, energy efficiency, and security. Buterin (2016) reviewed the opportunities and challenges of private and consortium blockchains, introducing the foundational ideas for Ethereum’s PoS approach \cite{buterin2016ethereum}. 

Kim (2020) outlined the structural changes and impacts of Ethereum 2.0 on the market, focusing on how staking services improve participation and decentralization \cite{kim2020ethereum}. Additionally, Brown-Cohen et al. (2019) explored the security barriers to implementing PoS protocols, providing insights into overcoming the "nothing-at-stake" problem \cite{brown2019formal}.

A recent study by Asif and Hassan (2023) delved into the energy consumption improvements brought by Ethereum 2.0's transition from Proof-of-Work to PoS, demonstrating the sustainability benefits of the new consensus mechanism \cite{asif2023shaping}. Furthermore, Dirdal and Bygdås (2024) focused on deploying private Ethereum 2.0 networks, presenting a more practical approach to managing PoS networks \cite{dirdal2024ethereum}.

These studies highlight key developments and solutions aimed at optimizing Ethereum’s PoS model, ensuring security, scalability, and sustainability.

Some studies focus on investigating the distribution of rewards among validators within staking pools. In \cite{9230398}, researchers have introduced reward-sharing schemes designed to facilitate equitable stakepool formation in collaborative projects with a substantial number of stakheolders. However, the primary objective of our research is to specifically address the operational mechanisms employed by two popular staking pools in resolving issues with staking. These issues notably include the considreable requirement for individuals to possess a substantial quantity of native cryptocurrency before they can partake as validators, rendering this process inaccessible for the majority of interested individuals seeking to engage in staking activities.

\section{Evolution of Ethereum}

The landscape of blockchain technology has been continually reshaped by the evolutionary journey of Ethereum, a pioneering platform that transcends traditional boundaries. In order to provide context about our analysis, we are going to provide the analysis of key milestones in Ethereum's development, focusing on the impact of the Beacon Chain and the Shanghai upgrade.

As Ethereum gained prominence, it encountered scalability challenges inherent in its original design. The surge in demand for decentralized applications and the limitations of the proof-of-work (PoW) consensus mechanism paved the way for Ethereum 2.0—an ambitious upgrade aiming to address these challenges.

Ethereum, since its inception in 2015, has been at the forefront of blockchain innovation, providing a decentralized platform for smart contracts and decentralized applications (DApps). Over the years, Ethereum has undergone several upgrades to enhance its scalability, security, and functionality. This section delves into the evolution of Ethereum, with a focus on two significant milestones—the introduction of the Beacon Chain and the Shanghai upgrade.

\subsection{Beacon Chain}

The Beacon Chain is the name of the original proof-of-stake blockchain that was launched in 2020. It was created to ensure the proof-of-stake consensus logic was sound and sustainable before enabling it on Ethereum Mainnet. Therefore, it ran alongside the original proof-of-work Ethereum. The Beacon Chain was a chain of 'empty' blocks, but switching off proof-of-work and switching on proof-of-stake on Ethereum required instructing the Beacon Chain to accept transaction data from execution clients, bundle them into blocks and then organize them into a blockchain using a proof-of-stake-based consensus mechanism. At the same moment, the original Ethereum clients turned off their mining, block propagation and consensus logic, handing that all over to the Beacon Chain. This event was known as The Merge.\cite{TheMerge} Once The Merge happened, there were no longer two blockchains. Instead, there was just one proof-of-stake Ethereum, which now requires two different clients per node. The Beacon Chain is now the consensus layer, a peer-to-peer network of consensus clients that handles block gossip and consensus logic, while the original clients form the execution layer, which is responsible for gossiping and executing transactions, and managing Ethereum's state. The two layers can communicate with one another using the Engine API.\cite{TheBeaconChain}

\subsection{Shanghai Upgrade}

Within the spectrum of Ethereum's evolution, the Shanghai upgrade stands out as a pivotal moment. This upgrade is strategically designed to enhance Ethereum's scalability and improve transaction throughput. By optimizing the Ethereum Virtual Machine (EVM) and introducing novel techniques such as statelessness, the Shanghai upgrade aims to streamline the network's efficiency.

Shanghai further reinforces Ethereum's commitment to continuous improvement, addressing bottlenecks and paving the way for a more responsive and scalable ecosystem. The synergy between the Beacon Chain and the Shanghai upgrade represents a holistic approach to Ethereum's maturation, fostering adaptability and resilience.

\section{Lido: A Comprehensive Analysis}
\subsection{Introduction}

Cryptocurrencies have witnessed remarkable growth and innovation in recent years, with Ethereum being at the forefront of this revolution. Ethereum's transition to a proof-of-stake (PoS) consensus mechanism with the launch of its beacon chain has opened up new possibilities for staking and earning rewards. In this paper, we delve into Lido, a prominent third-party staking solution that plays a pivotal role in Ethereum's PoS ecosystem. Lido has garnered attention for its unique mechanisms, tokenomics, and its contribution to Ethereum's liquid staking landscape.

\subsection{Mechanism and Working}

Lido operates as a liquid staking solution for Ethereum’s beacon chain\cite{9732095}. When users stake their Ethereum (ETH) on the beacon chain through Lido, they receive stETH tokens in return. These stETH tokens are designed to mirror the price of ETH almost 1:1, although there might be a slight gap. This mechanism allows users to stake their ETH while maintaining liquidity, as the stETH tokens can be traded or used in DeFi protocols. The staking process also enables users to earn rewards in the form of a certain percentage of the Annual Percentage Rate (APR) through Miner Extractable Value (MEV). Staking validators on Lido earn 5\% of the staking rewards, and another 5\% goes to the Lido DAO treasury. This distribution model incentivizes both validators for their service and the Lido DAO for maintaining and improving the protocol.

Decentralized Validator Technology (DVT) is a key component of Lido’s operations. DVT is a system that allows for the decentralization of staking pools. In traditional staking models, users delegate their tokens to a single validator, which can lead to centrailzation. However, with DVT, users’ stakes are distributed across multiple validators. This not only enhances the security of the protocol by reducing the risk associated with a single validator but also promotes a more decentralized and robust network. In the context of Lido, DVT allows for the distribution of staked ETH across multiple validators, thereby enhancing the security and decentralization of the staking process. This technology is crucial for maintaining Lido’s position as a leading liquid staking solution in the Ethereum ecosystem.

Miner Extractable Value (MEV) is a measure of the profit a miner can make through their ability to arbitrarily include, exclude, or re-order transactions within the blocks they produce. In the context of Lido, MEV is one of the ways users can earn rewards by staking their ETH. The APR for staking on Lido is currently at 3.8\%, which is the yield users can expect to earn from their staked ETH. This yield is a combination of staking rewards and profits from MEV. It’s important to note that while the stETH tokens mirror the price of ETH and can be traded or used in DeFi protocols, they also represent the staked ETH and the rewards earned, thereby allowing users to earn yield on their staked ETH.

\subsection{Tokenomics}

Lido's tokenomics are a testament to its commitment to decentralization. In addition to stETH, Lido has another native token called LDO, which was launched in December 2020. LDO serves as a governance token, allowing holders to participate in the decision-making processes of Lido DAO.

\begin{figure}[h]
\includegraphics[width=8cm]{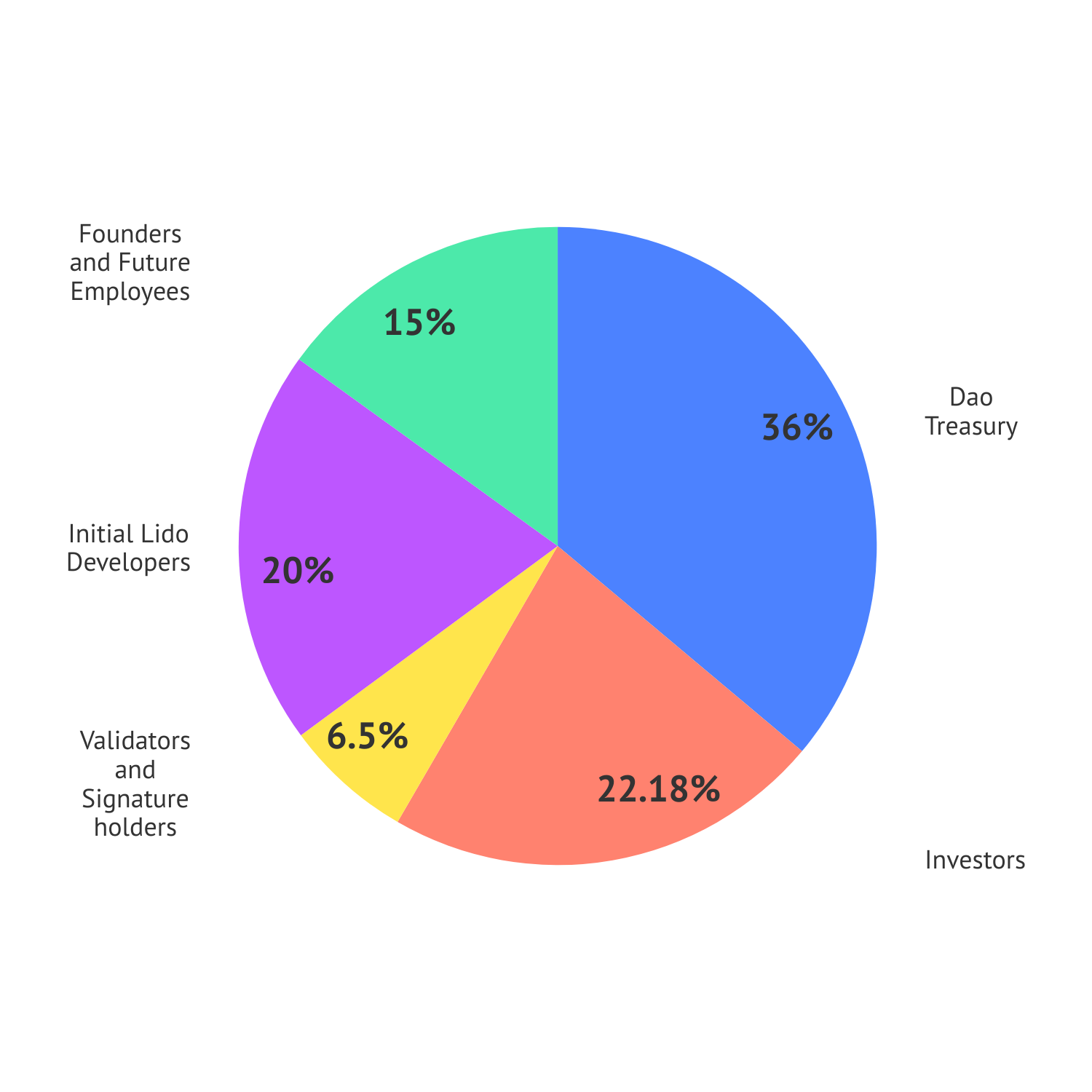}
\caption{Lido's initial distribution of tokens.}
\label{fig:yourlabel}
\end{figure}

\subsection{Liquid Staking and Beyond}

Lido's contribution to Ethereum's ecosystem goes beyond staking. It is recognized as a liquid staking protocol, allowing users to stake their ETH while maintaining liquidity. This feature enhances the flexibility and utility of staked assets, unlocking new possibilities for DeFi applications and lending.

Lido currently ranks \#8 on cryptocurrency charts by market capitalization, underscoring its importance and relevance in the crypto space.

\subsection{Challenges and Future Prospects}

While Lido has demonstrated remarkable performance, it is not without challenges. Two potential bug reports have been identified, but they have not caused significant damage to the protocol. These instances highlight the need for ongoing vigilance and security audits to safeguard the network.

As Ethereum's ecosystem continues to evolve, Lido's role in liquid staking and decentralized validation technology (DVT) is poised to become even more critical. StETH holders can actively participate in shaping the future of Lido through governance decisions, ensuring the protocol's adaptability and sustainability.

In conclusion, Lido stands as a key player in Ethereum's transition to PoS, offering an innovative liquid staking solution and a unique revenue-sharing model. With a robust tokenomics structure and active community involvement, Lido is well-positioned to play a pivotal role in the decentralized finance landscape.

\section{EigenLayer: The Restaking Collective}

\subsection{EigenLayer's Innovative Approach}
EigenLayer addresses the structural issues of Ethereum, including the lack of trust and security for AVSs, by establishing an open marketplace for decentralized security.

This solution addresses the issues mentioned above, including:
 
\begin{itemize}
  \item Mitigation of Captial Cost: One of the benefits of EigenLayer's approach is that it helps to mitigate capital costs. Instead of Ethereum L1 and AVS competing for limited capital resources, they can now cooperate and share their resources. ETH stakers can now double their profits by stake on both the Beacon chain and the AVS. Furthermore, AVS are no longer need to compelled to offer an aggressive staking pool during the initial bootstrapping phase, which reduces the overall cost burden. A graphic illustration is as Figure \ref{fig: eigenLayer_)omparing}.

\begin{figure*}[!t]
\centering
\includegraphics[width=0.7\textwidth]{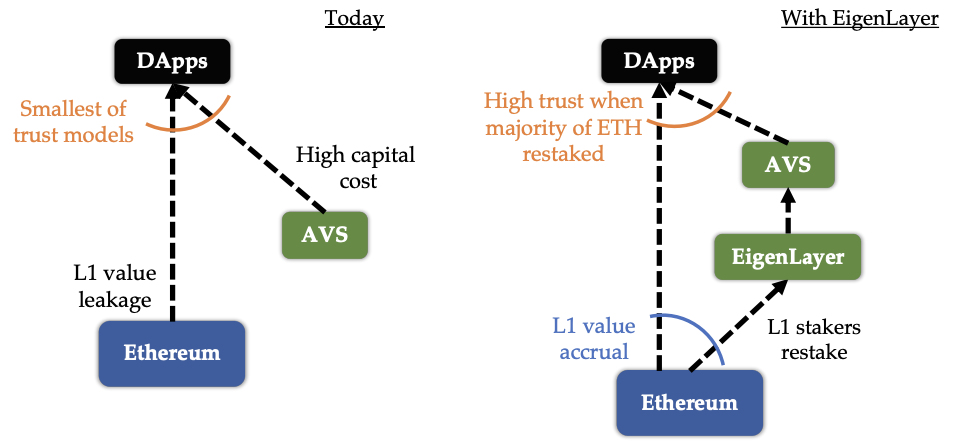}
\caption{Comparing the ecosystem of actively validated services today and with EigenLayer.}
\label{fig: eigenLayer_comparing}
\end{figure*}
  \item Fragmented Security: EigenLayer provides a new mechanism for pooled security. The Figure \ref{fig: Pooled security of EigenLayer} shows that \$13B is injected into the Ethereum ecosystem in both the left and right scenarios. However, in the left scenario where EigenLayer is not implemented, the cost of corruption (CoC) of AVS is \$1B, whereas, in the right scenario with EigenLayer, the CoC is increased to \$13B because of the improved security mechanisms.

\begin{figure*}[!t]
\centering
\includegraphics[width=0.7\textwidth]{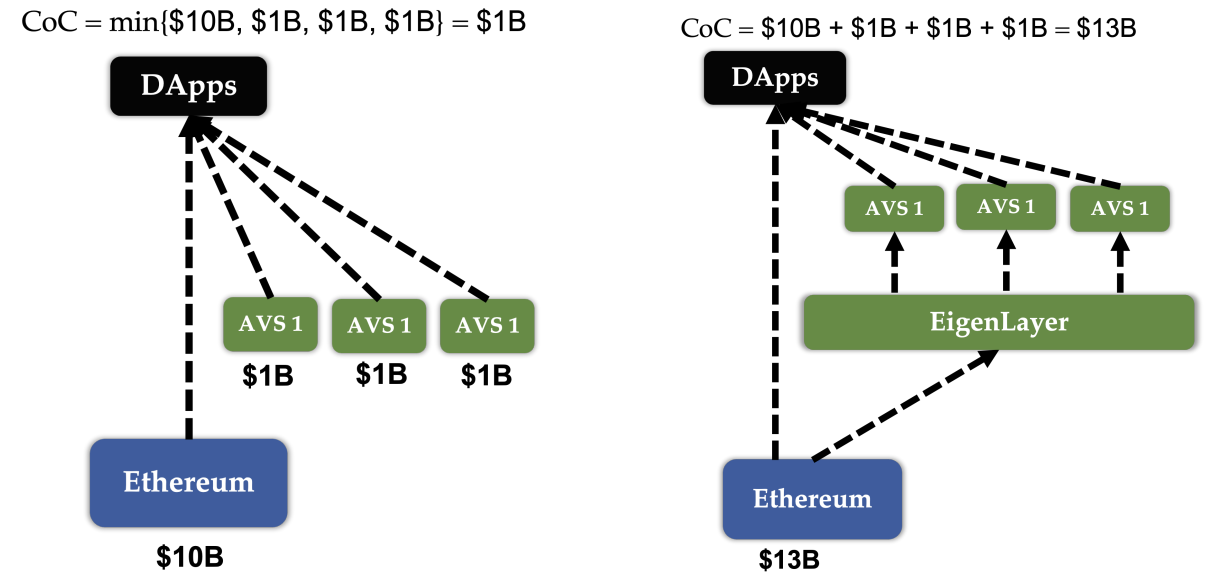}
\caption{Pooled security of EigenLayer.}
\label{fig: pooled_security_of_eigenLayer}
\end{figure*}

\end{itemize}

\subsection{Security and Efficiency}
EigenLayer introduces two novel ideas, pooled security via restaking and free-market governance\cite{EigenLayer_Whitepaper}, which serve to extend the security of Ethereum to any system and to eliminate the inefficiencies of existing rigid governance structures:

\textbf{1. Pooled security via restaking. }

EigenLayer provides a new mechanism for pooled security that allows modules to be secured by restaked ETH instead of their own tokens. Specifically, Ethereum validators can set their beacon chain withdrawal credentials to the EigenLayer smart contracts, and opt into new modules built on EigenLayer. The validators downloads and runs any additional node software required by the modules. Modules can then be able to impose additional slashing conditions on the staked ETH of validators who opted into the module. We call this mechanism restaking. In return, validators earn extra profit by providing security and validation services for the modules they choose. When coupled with onchain verifiable slashing mechanism, this restaking mechanism enables a deep transfer of cryptoeconomic security. For example, if the module is a Data Availability layer, restakers from EigenLayer will receive payment as long as the data is stored by the module.In return, restakers are subject to slashing conditions that are exercised through proof-of-custody. Restaking greatly expands the scope of blockchain applications that can integrate security. As a result, EigenLayer continues to unlock innovation beyond smart contract-based DApps powered by Ethereum, to virtual machines, consensus protocols, and middleware. An AVS (Active Authentication Service) with an on-chain slashing contract can be protected by EigenLayer.

\textbf{2. Open marketplace. }

EigenLayer provides an open market mechanism which governs how its pooled security is supplied by validators and consumed by AVSs. EigenLayer creates a marketplace in which validators can choose whether to opt in or out of each module built on EigenLayer. The various modules will need to sufficiently incentivize validators to allocate restaked ETH to their module and validators will help determine which modules are worthwhile to assign this additional pooled security given the potential for additional slashing. The opt-in dynamics of EigenLayer have two important benefits: (1) the stable, conservative governance of the core blockchain is complemented with a fast and efficient, free-market governance structure for launching new auxiliary capabilities; and (2) opt-in validation makes it possible for new blockchain modules to exploit heterogeneous resources among validators, resulting in better tuned trade-offs of security and performance.

\subsection{Delegation in EigenLayer}

Delegation in EigenLayer is the process by which stakers can delegate their stake to other entities, known as operators. These operators run EigenLayer nodes and participate in various services, such as validating Ethereum blocks and providing attestation services to other protocols.

Stakers delegate their stake to operators by depositing their ETH into an EigenLayer delegation contract. The delegation contract then creates a delegation pool, which is a collection of ETH that has been delegated to a particular operator.

Once a staker has delegated their stake to an operator, the operator is responsible for managing the delegated ETH and participating in the appropriate services. The operator will earn staking rewards on behalf of the stakers, and will distribute these rewards to the stakers at regular intervals.

Stakers can choose to delegate their stake to any operator that they trust. Operators can compete for delegators by offering competitive fees and services. Stakers can also diversify their risk by delegating to multiple operators.

Delegation in EigenLayer is a powerful tool that can help to create a more decentralized, secure, and accessible Ethereum ecosystem. By making it easier for stakers to participate in the network, delegation can lead to a more robust and resilient network. Additionally, delegation can help to reduce the risk of centralization by distributing the power of the network among more participants.

\subsection{Slashing Mechanism}
Cryptoeconomic security quantifies the cost that an adversary must bear in order to cause a protocol to lose a desired security property. This is referred to as the Cost-of-Corruption (CoC)\cite{The_cryptoeconomics_of_slashing}. When CoC is much greater than any potential Profit-from-Corruption (PfC), we say that the system has robust security. Cryptoeconomic security stands in contrast with systems that provide majority-trust security guarantees which only hold under the assumption that at least a threshold percentage of operators are altruistic and will act honestly. A core idea of EigenLayer is to provision cryptoeconomic security through various slashing mechanisms which levy a high cost of corruption.

A key function of the EigenLayer smart contracts is to hold the withdrawal credentials of Ethereum Proof-of-Stake (PoS) stakers. If a staker who is restaked on EigenLayer is proven to have behaved adversarially while participating in an AVS, then that staker’s ETH will be subject to slashing and are frozen, that is, prevented from further participation on any AVS on EigenLayer. Since the withdrawal address of the staker is set to the EigenLayer contracts, when the staker withdraws their ETH from participation in Ethereum consensus through EigenLayer, the withdrawn ETH will be slashed according to the on-chain slashing contract of the AVS.

\subsection{Impact of EigenLayer}
In any distributed system, there is a natural tendency for stake and validator nodes to centralize. This is because it becomes more efficient and cost-effective to run a single, large validator node than multiple smaller nodes. However, many AVSs rely on having their stake and validator nodes be as decentralized as possible. For example, in a threshold encryption system for on-chain privacy, it is important that the set of threshold key holders be decentralized so that no one entity can know the actual content before the stipulated period.

EigenLayer is a Layer-2 scaling solution for Ethereum that enables AVSs to explicitly specify that the stake/nodes participating in their validation tasks must be part of a decentralized quorum. This means that AVSs can specify that only Ethereum home validators can participate in their tasks. This permissionless ability to specify who can participate in the validation tasks of an AVS is equivalent to programming decentralization.

At an abstract level, EigenLayer enables a marketplace for AVSs to buy decentralization. As more and more AVSs specify that only home validators can participate in their tasks on EigenLayer, this makes it more profitable to run home validator nodes on Ethereum, which incentivizes decentralization.

In summary, EigenLayer is a powerful tool that can be used to promote decentralization in distributed systems. By enabling AVSs to programmatically specify the decentralization requirements for their validation tasks, EigenLayer creates a marketplace for decentralization that incentivizes the growth and development of the Ethereum home validator network.

\section{Solutions}

\subsection{High barrier to become a validator}

How can Eigenlayer solve this problem? EigenLayer is a restaking collective for Ethereum that enables consensus layer Ether (ETH) stakers to validate new software modules built on top of the Ethereum ecosystem.\cite{EigenlayerAnalysisReport} Its restaking feature enables staked ETH to be used as cryptoeconomic security for protocols other than Ethereum, in exchange for protocol fees and rewards. Restaking is available for both natively staked ETH and liquid staked tokens like stETH, rETH, cbETH, and LsETH.\cite{EigenLayer_Whitepaper} EigenLayer also provides developers with access to the Ethereum staked capital base and decentralized validator set, which can make previously impossible mechanism designs possible.

EigenLayer’s restaking feature could help reduce the high barrier to become a validator in Ethereum by enabling validators to earn fees on additional services and restake their assets on other emerging networks. EigenLayer also allows validators to participate in new consensus protocols which have low latency and high throughput..

Unlike Eigenlayer, Lido provides another approach for solving this problem. Lido enables users to stake their ETH without having to run their own validator node. Instead, Lido pools the ETH of multiple users and stakes it on their behalf. In exchange, users receive stETH, a liquid representation of their staked ETH, which can be traded or used in other DeFi protocols.\cite{WhatIsLido2021}

Lido’s restaking feature enables staked ETH to be used as cryptoeconomic security for protocols other than Ethereum, in exchange for protocol fees and rewards. Restaking is available for both natively staked ETH and liquid staked tokens like stETH, rETH, cbETH, and LsETH.\cite{WhatisLidoLDO} Lido also allows validators to participate in new consensus protocols which have low latency and high throughput.

Lido’s restaking feature could help reduce the high barrier to become a validator in Ethereum by enabling validators to earn fees on additional services and restake their assets on other emerging networks. Lido is Ethereum’s largest liquid staking project, and it is three times larger than the next-largest validator, Coinbase. Lido’s restaking feature enables staked ETH to be used as cryptoeconomic security for protocols other than Ethereum, in exchange for protocol fees and rewards. Lido also allows validators to participate in new consensus protocols which have low latency and high throughput.

In conclusion, Lido solves the high barrier to become a validator in Ethereum by enabling users to stake their ETH without having to run their own validator node, and EigenLayer solves it by enabling users to restake their assets on other emerging networks and earn fees on additional services. 

    \subsection{The staked Ether is currently locked and cannot be withdrawn immediately for certain amount of time }

        In addressing the second issue pertaining to the staked ETH or Ether lock-in period and its implications on user accessibility, it is imperative to delve into the intricate details surrounding this phenomenon. The predicament arises from the inherent nature of the staked ETH, which undergoes a temporal restriction, rendering it inaccessible for a predetermined duration. This issue not only necessitates a thorough examination but also underscores the significance of understanding the evolving landscape of Ethereum's protocol upgrades.
        
         A noteworthy observation stems from the temporal constraints imposed on the withdrawal of staked ETH, a subject that was previously elucidated in an article on ethereum.org. According to the initial discourse, a lock-in period of 6-12 months post the protocol merge was anticipated. However, it is noteworthy to mention that the article underwent subsequent modifications, ultimately leading to its update or deletion within a few months. This dynamic nature of information dissemination within the Ethereum ecosystem underscores the need for a meticulous exploration of historical data to discern the underlying trends.
                  \begin{figure*}[!t]
    \centering
    \includegraphics[width=0.7\textwidth]{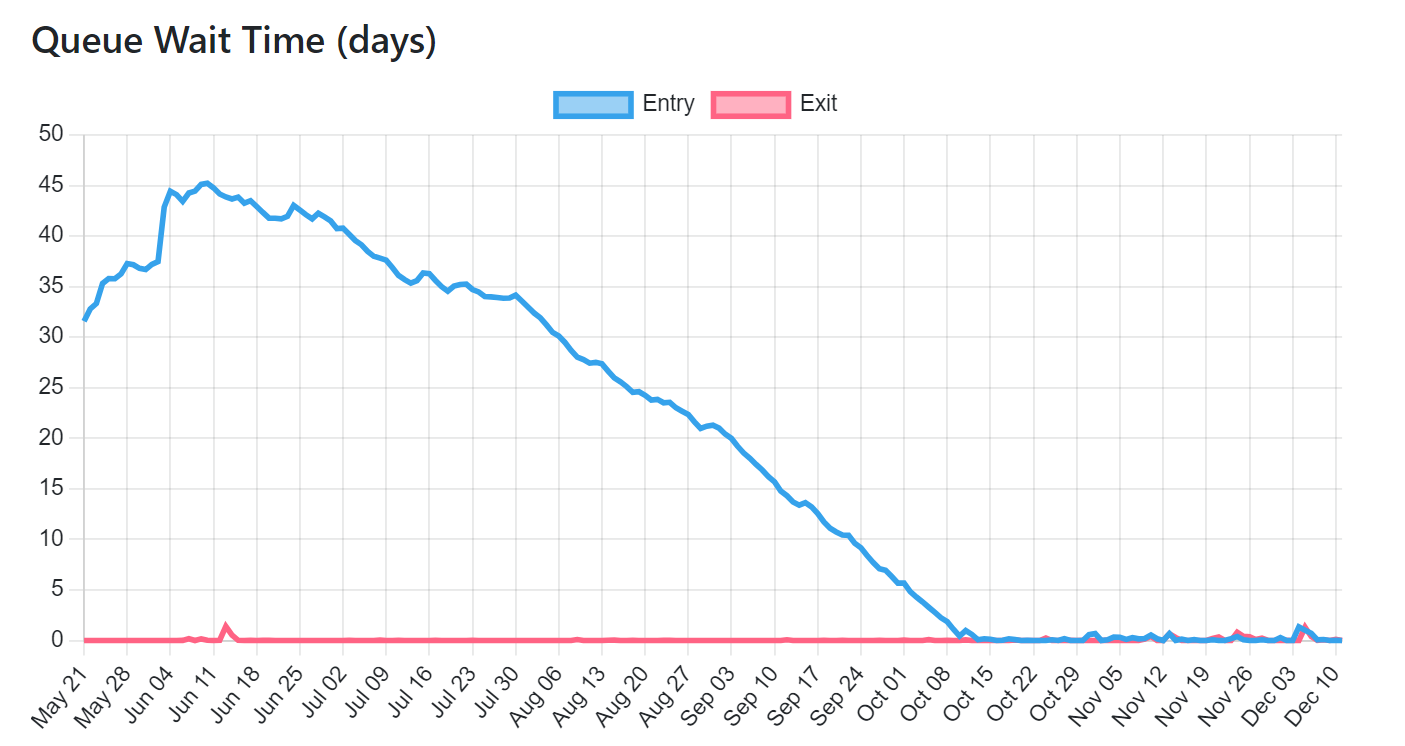}
    \caption{Historical entry/exit wait time}
    \label{fig: waittime}
\end{figure*}

        To gather insights into this intricate matter, we meticulously collected data from diverse sources, including the beaconchain API, ultrasound.money API v2, and Ethereum node data from Infura and Etherscan. The amalgamation of these datasets has facilitated a comprehensive analysis of the Entry and Exit queue wait times, as illustrated in the accompanying graph. The graph delineates the Entry Queue time represented by the discernible blue line and the Exit Queue time depicted in a conspicuous red color.
        
         Before delving further into our findings, it is crucial to acknowledge the commencement of our data collection, which initiated on May 21st. However, the Ethereum network underwent the Shanghai Upgrade on April 12th, leading to a temporal hiatus of more than 35 days in our dataset. Despite this challenge, our commitment to delivering robust insights propelled us to explore alternative avenues for statistical validation.

        In our quest for additional data points, we ventured beyond conventional sources and scrutinized social media platforms such as Twitter and Reddit. The rationale behind this approach was to tap into the narratives of prominent entities within the blockchain space, including whales and industry influencers. Our findings from these unconventional sources revealed a nuanced perspective, indicating that, in the aftermath of the protocol merge, the withdrawal queue extended beyond 20 days. This temporal lag, as observed in real-world scenarios, contrasts with the graphical representation in our dataset. While the graph may not entirely mirror the magnitude of the withdrawal queue, the dissonance is noteworthy. It prompts a critical examination of the significance of real-time observations and experiential insights, which often transcend the confines of quantitative data representation.
\begin{algorithm}
\caption{Calculate Wait Time}
\KwIn{active, queue}
\KwData{scaling, epochChurn, dayChurn}
\KwOut{waitTime, currChurn, aveChurn, churnTimeDays}
\BlankLine

The purpose of this algorithm is to estimate the time a validator has to wait before entering or exiting the Ethereum 2.0 network. It takes into account the number of active validators and the size of the entry or exit queue.

$currChurn \leftarrow 9$\;
$churnTimeDays \leftarrow churnFactor \leftarrow 0$\;

\For{$i$ \textbf{from} 0 \textbf{to} \texttt{LENGTH(scaling) - 1}}{
    \If{active > scaling[i]}{
        $currChurn \leftarrow$ epochChurn[i]\;
    }
    
    \If{i < LENGTH(scaling) - 1 \textbf{and} active >= scaling[i] \textbf{and} active < scaling[i + 1]}{
        $j \leftarrow i$\;
        $remain \leftarrow$ queue\;

        \While{remain > 0}{
            \If{i === j}{
                $churnTimeDays \leftarrow churnTimeDays + \min\left(\frac{remain}{\text{dayChurn}[j]}, \frac{(scaling[j + 1] - \text{active})}{\text{dayChurn}[j]}\right)$\;
                $churnFactor \leftarrow churnFactor + \min\left(\text{remain}, (scaling[j + 1] - \text{active})\right) \times \text{epochChurn}[j]$\;
                $remain \leftarrow 0$\;
            }
            \Else{
                $churnTimeDays \leftarrow churnTimeDays + \min\left(\frac{remain}{\text{dayChurn}[j]}, \frac{(scaling[j + 1] - \text{scaling[j]})}{\text{dayChurn}[j]}\right)$\;
                $churnFactor \leftarrow churnFactor + \min\left(\text{remain}, (scaling[j + 1] - \text{scaling[j]})\right) \times \text{epochChurn}[j]$\;
                $remain \leftarrow remain - \min\left(\text{remain}, (scaling[j + 1] - \text{scaling[j]})\right)$\;
            }
            
            $j \leftarrow j + 1$\;
        }
    }
}

$aveChurn \leftarrow$ \textbf{if} queue > 0 \textbf{then} $\frac{\text{ROUND}(churnFactor / \text{queue} \times 100)}{100}$ \textbf{else} currChurn\;

$waitSecs \leftarrow \text{ROUND}(churnTimeDays \times 86400)$\;
$waitDays \leftarrow \text{FLOOR}(waitSecs / 86400)$\;

$waitTime \leftarrow$ \textbf{if} waitDays > 0 \textbf{then} `$waitDays$ \textbf{day(s)}" \textbf{else} $\text{FLOOR}(waitSecs / 3600)$ \textbf{hour(s)}, $\text{ROUND}(((waitSecs \% 3600) / 60))$ \textbf{minute(s)}"$\textbf{fi}$\;

\texttt{PRINT "Churn Time Days:", churnTimeDays}\;
\texttt{PRINT "Current Churn:", currChurn}\;
\texttt{PRINT "Average Churn:", aveChurn}\;

\KwRet{waitTime, currChurn, aveChurn, churnTimeDays}\;

\end{algorithm}

        In conclusion, our comprehensive investigation into the second issue surrounding the staked ETH lock-in period provides valuable insights into the multifaceted nature of this phenomenon. While certain aspects of the challenge have seen resolution, as evidenced by the notably short exit queue wait times—contrary to Ethereum's initial prognostication of extended delays—the overarching issue persists. The crux of the matter lies in the inherent constraint imposed on staked Ether, rendering it immobilized and precluding its utilization for alternative purposes. The observed brevity in exit queue wait times stands as a testament to the adaptability of the Ethereum network and its capacity to navigate challenges effectively. However, it is imperative to underscore that the core predicament persists in the form of staked Ether remaining inaccessible for other applications once committed to the staking protocol. This nuanced understanding contributes to the ongoing discourse surrounding Ethereum's protocol upgrades, shedding light on both resolved and lingering aspects of the staking infrastructure.

In essence, while strides have been made in mitigating certain facets of the staked ETH lock-in challenge, the overarching issue remains an integral part of the decentralized ecosystem. This academic exploration aims to foster a nuanced comprehension of the dynamic interplay between protocol upgrades and the evolving landscape of blockchain technology, emphasizing the ongoing nature of research required to address the intricacies inherent in decentralized systems.
\subsubsection*{\textbf{Explanation of the Algorithm}}

In the algorithm, the churn calculation loop plays a crucial role in estimating the Ethereum validator entry and exit queue times. Let's explore this loop and the associated variables in detail.

\begin{itemize}
    \item \textbf{Churn Calculation Loop:}

The loop is designed to adapt the churn calculations based on the number of active validators and the scaling ranges. It iterates through these ranges, determining the appropriate churn values for the given conditions.

\item \textbf{Variables:}

\begin{itemize}
    \item \texttt{i:} Loop variable representing the index of the scaling array.
    \item \texttt{j:} Helper variable for nested calculations within the loop.
    \item \texttt{remain:} The remaining size of the entry or exit queue being processed.
\end{itemize}

\item \textbf{Active Validators:}

The variable \texttt{active} represents the number of active validators in the Ethereum network. The loop adjusts the churn calculations based on the current number of active validators.

\item \textbf{Scaling, EpochChurn, DayChurn:}

These arrays provide predefined values for different scaling ranges, corresponding epoch churn values, and day churn values. The loop references these arrays to determine the appropriate values for churn calculations.

\item \textbf{Churn Time and Factor:}

\begin{itemize}
    \item \texttt{churnTimeDays:} Cumulative churn time in days.
    \item \texttt{churnFactor:} Cumulative churn factor considering the queue size.
\end{itemize}
The loop calculates these values dynamically, adjusting them based on the specific conditions within each scaling range.

\item  \textbf{Average Churn:}

After processing the entire queue, the algorithm calculates the average churn (\texttt{aveChurn}) based on the churn factor and the size of the queue. If the queue is empty, it defaults to the current churn value.

\item \textbf{Output Printing:}

The algorithm prints essential information related to churn, including churn time in days, current churn, and average churn. These outputs provide insights into the dynamics of the Ethereum network.

\item  \textbf{Return Statement:}

The loop contributes to the overall calculation, and the algorithm returns key metrics, including the wait time, current churn, average churn, and churn time in days.

This thorough explanation highlights how the churn calculation loop and associated variables contribute to the estimation of validator entry and exit queue times in the Ethereum network.

\end{itemize}

\subsection{Small number of entities could control a large majority of the staked ETH}

EigenLayer is a middleware built on the Ethereum network that aims to commoditize decentralized trust. It allows protocols that integrate with it to leverage Ethereum’s highly secure trust network without needing to establish its own validator set, offering an off-chain data availability option to Layer 2s to further reduce costs. EigenLayer’s restaking mechanism leverages pooled security and free-market governance to extend Ethereum’s base layer security to essentially any protocol built on top of it, irrespective of its composition.

EigenLayer offers an alternative off-chain data availability option to Layer 2s, enabling them to save on the data management cost while providing a similar level of security. Users benefit from increased capital efficiency by having the option to restake their ETH, in turn receiving more staking rewards.

Overall, EigenLayer provides a solution to the challenge of a small number of entities controlling a large majority of the staked ETH in Ethereum by offering an alternative off-chain data availability option to Layer 2s, enabling them to save on the data management cost while providing a similar level of security. Additionally, users benefit from increased capital efficiency by having the option to restake their ETH, in turn receiving more staking rewards.

Lido is a decentralized autonomous organization (DAO) that provides liquid staking services for Ethereum. It allows users to stake their ETH and receive stETH in return, which can be used for other purposes while still earning staking rewards. Lido’s liquid staking service is designed to solve the problem of illiquidity and centralization in staking. By providing a liquid staking solution, Lido enables users to trade staked ETH on decentralized exchanges, which increases liquidity and reduces the risk of centralization. Additionally, Lido is a non-custodial solution, which means that users retain control over their staked ETH at all times. This helps to prevent a small number of entities from controlling a large majority of the staked ETH, which is a major challenge in Ethereum staking.

\section{Conclusion}
In wrapping up our exploration into the intricacies of Ethereum's Proof of Stake (PoS) consensus mechanism and the potential remedies offered by EigenLayer and Lido, it's evident that these solutions bring a refreshing breeze to the blockchain space. The challenges faced by PoS have met their match in the form of EigenLayer, a middleware that cleverly taps into Ethereum's robust trust network without burdening itself with a standalone validator set. This off-chain data availability option opens new doors, particularly for Layer 2s, promising cost reductions and enhanced efficiency.

Enter Lido, the beacon of simplicity in the staking world. By allowing users to stake any amount of ETH and earn rewards on the Beacon Chain, Lido eliminates the daunting 32 ETH entry barrier for running a validator node. Not only does it lower the financial threshold, but it also liberates users from the intricacies of technical maintenance associated with managing a staking node.

As we reflect on EigenLayer and Lido's contributions, it becomes clear that they're not just solutions; they're catalysts for a more inclusive and streamlined Ethereum PoS ecosystem. The challenges we've dissected are dynamic, ever-evolving entities, and these innovative approaches signal a promising future. In the grand symphony of technological advancements and community-driven initiatives, EigenLayer and Lido have taken the stage, playing a key role in harmonizing the debate and discussions surrounding Ethereum's PoS landscape.

\vspace{12pt}

\end{document}